%% file: main.tex
\documentclass[conference]{IEEEtran}
\IEEEoverridecommandlockouts
\usepackage{pgfplots}
\pgfplotsset{compat=1.18}
\usepackage{amsfonts}
\usepackage{xurl}
\usepackage{cite}
\usepackage{url}
\usepackage{amsmath,amssymb,amsfonts}
\usepackage{algorithmic}
\usepackage{graphicx}
\usepackage{tikz} 
\usepackage{textcomp}
\usepackage{xcolor}

\usepackage[colorlinks=true,linkcolor=blue,citecolor=blue,urlcolor=blue]{hyperref}
\def\BibTeX{{\rm B\kern-.05em{\sc i\kern-.025em b}\kern-.08em
    T\kern-.1667em\lower.7ex\hbox{E}\kern-.125emX}}
\begin{document}

\title{Optimal Low Emission Zones scheduling as an example of transport policy backcasting \\
}

\author{\IEEEauthorblockN{Asmae ALAMI}
\IEEEauthorblockA{
\textit{IFP Energies Nouvelles}\\
Rueil-Malmaison, France \\
asmae.alami-ouazzani@ifpen.fr}
\and
\IEEEauthorblockN{Vinith LAKSHMANAN}
\IEEEauthorblockA{
\textit{IFP Energies Nouvelles}\\
Rueil-Malmaison, France \\
vinith-kumar.lakshmanan@ifpen.fr}
\and
\IEEEauthorblockN{Antonio SCIARRETTA}
\IEEEauthorblockA{
\textit{IFP Energies Nouvelles}\\
Rueil-Malmaison, France \\
antonio.sciarretta@ifpen.fr}
}

\maketitle

\begin{abstract}
This study presents a backcasting approach that considers the passenger car fleet dynamics to determine optimal policy roadmaps in transport systems. As opposed to the scenario-based approach, backcasting sets emission reduction targets first, then identifies policies that meet the constraint. The policy is the implementation of Low Emission Zones (LEZs), in the Île-de-France region as a case study. The aim is to minimize the number of scrapped vehicles due to LEZs under CO$_2$ emission targets and to deduce an interdiction schedule of polluting vehicles by 2050. 
To explore potential solutions, we use a genetic algorithm that provides a first insight into optimal policy pathways. 

\end{abstract}


\section{Introduction}
Air pollution is responsible 
for exposing populations to dangerous and toxic emissions significantly impacting their health 
sometimes even leading to death \cite{pollution-mortality}. 
Greenhouse gas emissions (GHG) significantly contribute to global warming, 
a key topic that remains at the heart of public debate in policy-making and campaigning across Europe\cite{climate-change-politics-poland}. The  European Union has set an ambitious goal of achieving carbon neutrality by 2050, requiring a massive reduction of transportation sector emissions, as it remains one of the most polluting sectors and one of the largest contributors to GHG emissions\cite{StatistaGHG}.


Many cities have been exploring numerous strategies to reduce vehicle pollutant emissions and decrease GHG emissions, mainly by reducing the use of private cars and lowering congestion. Congestion pricing in urban areas introduces road tolls to regulate traffic-flow and discourage unnecessary trips during peak hours. 
In a study on Lyon (France) \cite{urban_tolls_lyon}, it has been shown that urban tolls generate revenue that can be reinvested in public transport and sustainable infrastructure, thereby improving accessibility and reducing CO$_2$ emissions. Another study \cite{urban-tolls_vs_lez} highlights that urban tolls are more effective when combined with public transport improvements while acceptance remains a challenge. 

Low Emission Zones (LEZ) have significantly spread across Europe\cite{ArvalLEZ} and became one of the most popular measures in terms of reducing GHG emissions and improving public health \cite{health-germany}. Although LEZs have shown effectiveness in reducing PM$_{10}$ and NO$_2$ concentrations in cities like Munich or Berlin \cite{Critical-review-lez}, the effectiveness in other areas such as London has shown to be minimal \cite{improved-ULEZ?}. In Madrid, a study \cite{scrapping-lez-madrid} has shown that the introduction of even a small LEZ entails scrapping old vehicles, another study \cite{Espagne-ZFE-et-ZEZ} highlights that LEZs encourage the adoption of alternative-fuel vehicles, all of which inevitably improves air quality. In spite of  LEZs having great potential in sustainable mobility, their implementation can lead to a decline in people's well-being \cite{well-being}. This is mainly due to restrictions on individual mobility and common requirement to replace forbidden vehicles with costly alternatives, often with little or no government support\cite{ReflexScienceLEZ}. Therefore, it is important that policymakers also conduct thorough studies to assess LEZs impacts on the population to make an informed decision. 

More generally, it is vital to assess the effectiveness of different policies to tackle the reduction of air pollutants, GHG, and guide the transport sector's carbon neutrality. 
A large body of litterature \cite{article-me}, \cite{SMA_human_air_pollution}, \cite{SMApollution_impacts}, uses agent-based simulation for example, to forecast the performance of the decisions to be made in terms of improvement of air quality and/or reduction of GHG emissions via prospective scenarios. 
Either by selecting the best scenario among the designed ones or trying to explore a larger scale of possibilities via a surrogate model (Forecasting) and then complete with an optimization process (Backcasting), the agent-based approach shows its limitations as many uncertainties need to be quantified to approve of the results.

In this paper, we are pursuing the research conducted by Lakshmanan \textit{et al.}\cite{misc-Vinith} to explore a new application of the backcasting paradigm on LEZs with an additionnal focus on its acceptability among users by considering supplementary indicators. The subsystem used for this paradigm describes the evolution of passenger vehicle fleet and its impact on GHG emissions as well as impact on individuals behaviour in response to LEZ policies. The decision variable corresponds to the detailed schedule of vehicles' age to forbid from present to 2050 within a certain perimeter/region. This paper is organised as follows : in Section~\ref{Model} we introduce the new revised model that takes into account LEZs then the formulation of the optimal control problem. Section~\ref{case-study} presents our case study that covers the Parisian Region (Île-de-France). We will present our results in Section~\ref{results} and discuss optimal schedules. And finally in Section~\ref{conclusion} we will conclude and talk about perspectives for future work.

\section{Backcasting model}
\label{Model}

\subsection{Vehicle fleet model }
\label{Model_1}

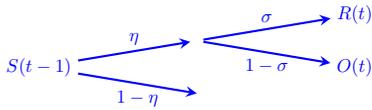
\begin{figure}[!h]
    \centering
    \input{survival}
    \caption{Schematic representation of the relationship between $S$, $O$ and $R$}
    \label{fig-survival-rate_riddance}
\end{figure}

 We use the time variable $t \in [\![0,T]\!]$ starting from present until year $2050$ with $T$ representing the total number of years between both periods, and consider that the vehicle stock $S(t)$ is composed of 2 types of vehicles : thermal ($v=1$) and electric ($v=2$), and that their ages range from $a=0$ (new vehicle) to $a=A=30$. In this model, we consider a large study area (cf. section~\ref{case-study}) that we divide into $Z$ complementary disjoint zones and introduce a disaggregation variable $z \in [\![1,Z]\!]$ which corresponds to one of the segmented zones. 
Vehicle-km (vkm) is a measure of the transport demand $G(t)$, which can be segmented into distinct zones, and the mileage $M(t)$ is the average distance per vehicle per year (km/veh/y). $N(t)$ is the number of new vehicles ($a=0$) needed to satisfy the total demand and is expressed as follows 
\begin{equation}
N_z(t) = \frac{G_z(t)}{M(t)} - \sum_{va} O_{vaz}(t) \label{new vehicles}
\end{equation}
with $O(t)$ being the total number of old vehicles ($a>0$) at year $t$, segmented into type, age and zone 
: $O_{vaz}(t)$. Because of LEZs, most vehicle owners will dispose of their vehicle if it has been or is close to being forbidden. Let $R(t)$ be the number of vehicles that were disposed of because of the LEZ and $\sigma$ the disposal ratio that depends on the LEZ schedule. Just as established in \cite{misc-Vinith}, 
the $\eta_a$ represents the survival rate of a vehicle from a year to the next, depending on its age $a$. Figure~\ref{fig-survival-rate_riddance} shows that among the previous available stock $S(t-1)$, the vehicles that survive, are either disposed of ($R(t)$) or become part of the old stock ($O(t)$). The vehicles that do not survive leave the fleet and are not quantified. 

The expressions of old and disposed vehicles are valid $\forall t \in [\![1,T]\!]~$ and are expressed as follows
\begin{equation}
    \centering
    \begin{aligned}
    O_{vaz}(t) &= 
        \eta_a (1-\sigma_{vaz}(t))S_{v,a-1,z}(t-1) \quad \forall a \in [\![1,A-1]\!]\\
    O_{vAz}(t) &= \eta_A (1-\sigma_{vAz}(t))[S_{v,A-1,z}(t-1)+S_{vAz}(t-1)]
    \label{old-stock}
    \end{aligned}
\end{equation}
\begin{equation}    
    \centering
    \begin{aligned} 
        R_{vaz}(t) &= \eta_a \sigma_{vaz}(t)S_{v,a-1,z}(t-1) \quad \forall a \in [\![1,A-1]\!]\\
        R_{vAz}(t) &= \eta_A \sigma_{vAz}(t)[S_{v,A-1,z}(t-1)+S_{vAz}(t-1)].
    \end{aligned}
    \label{riddance-stock}
\end{equation}

The disposal ratio $\sigma_{vaz}(t)$ depends on the LEZ and will be discussed in subsection~\ref{decision-variable}. In this paper, we assume that the only vehicles that can be disposed of 
 are thermal vehicles, which implies that $\sigma_{2az}(t) = 0 ~ 
 \forall t$ and $\forall z$. 
 
 LEZs also change the course of purchase of thermal vehicles. Indeed, if a LEZ forbids access to a certain vehicle, its owner is less likely to replace it with a new thermal vehicle. 
In this paper, we consider that if a vehicle has been denied access to because of a LEZ, then the probability of replacing it with a thermal vehicle is null. 
The number of new vehicles has also been disaggregated by type and is expressed as follows
\begin{subequations}
\begin{align}
        N_{1z}(t) &= (N_z(t) - \sum_a R_{1az}(t))P_{1z}(t) \label{new-vehicles-typeT} \\
        N_{2z}(t) &= \sum_a R_{1az}(t)+ (N_z(t) - \sum_a R_{1az}(t))P_{2z}(t). \label{new-vehicles-typeE}
\end{align}
\label{new_vehicles_type}
\end{subequations}
where $P_{1z}(t)$ and $P_{2z}(t)$ are the ratios of thermal and electric vehicle purchase. Expressions for $P_{vz}(t)$ as a function of fixed and variable costs will be introduced in subsection~\ref{decision-variable}.

The evolution of vehicle stock by vehicle type, age and zone at year $t$ is expressed as 
\begin{equation}
    S_{vaz}(t) = 
    \begin{cases}
        N_{vz}(t) \quad a=0 \\
        O_{vaz}(t) \quad a>0.
    \end{cases}
    \label{stock}
\end{equation}

From (\ref{stock}), CO$_2$ emissions are evaluated using emission factor $ \epsilon_{1a}$ (gCO$_2$/km) such that $ \epsilon_{2a}=0$ for electric vehicles. 
\begin{equation}
    E(t)=M(t)\sum_{a} \epsilon_{1a}(t)\sum_z S_{1az}(t).
    \label{emission_function}
\end{equation}

\subsection{Decision Variable and disposal function}
\label{decision-variable}
\subsubsection{Age restriction variable}

~

Let us introduce the age restriction variable $I\in \mathbb{N}^{T\times Z}$, such that $0\leq I_z(t) \leq A$. 
Each element $I_z(t)$ represents ages of vehicles and indicates from which age vehicles are forbidden. The initial value $I_z(0)$ is a fixed parameter on which we base the study and is set in table~\ref{tab:schedule}. If $I_z(t)=a$, all vehicles of age $a$ and more are forbidden in zone $z$ at time $t$. As only thermal vehicles are assumed to be forbidden, the subscript $v$ is dropped in $I$.
We impose that $I$ verifies certain constraints moving from a year to the next. Our first constraint consists of not allowing a forbidden vehicle to be authorized again.
 This means we impose that for the same zone 
 , $I_z(t) \leq I_z(t-1)+1 \quad \forall t \in [\![1,T]\!]$.
If no new restriction is applied moving from a year $t$ to the next, $I_z(t) = I_z(t-1)+1$.
The second constraint to impose is that the ban is reasonably restrictive. To illustrate this constraint, for each zone, we allow a maximum ban slope between consecutive years. The parameter in question is $D \in \mathbb{N}^Z$, and is intended to be set by decision-makers, but specific values were assigned to $D$ in Table~\ref{tab:schedule} to serve as a reference for this study. 
Having a fixed threshold for each zone, the second constraint would be 
    $I_z(t-1) - D_z \leq I_z(t) \quad \forall t \in [\![1,T]\!]$.
Both constraints depend on the previous value of $I$; thus, we introduce a decision variable $J\in \mathbb{Z}^{T\times Z}$, such that \begin{equation}
    I_z(t-1)-I_z(t) = J_z(t). \quad 
    \label{relation J et I}
\end{equation}
The resulting constraint for the decision variable is 
\begin{equation}
    -1 \leq J_z(t) \leq D_z, \quad 
    \label{constrJ}
\end{equation}
and the state variable $I$ is further constraint by 
\begin{equation}
    0 \leq I_z(t) \leq A. \quad 
    \label{constrI}
\end{equation}
The decision variable $J$ explicitly shows how many additional ages will be forbidden at time $t$ compared to ones forbidden at time $t-1$. 
This makes $J$ a suitable control variable for the optimal control problem and $I$ one of its state variables. 

In addition, $\Pi_{az}(t)$ is a variable that indicates if a vehicle of age $a$ is forbidden in zone $z$ at time $t$. It equals $1$ if the vehicle is forbidden or $0$ if it is not, and is defined as
\begin{equation}
    \Pi_{az}(t) = \mathbf{1}_{\{I_z(t)\leq a\}}= \begin{cases}
        1 \quad \text{if } I_z(t) \leq a \\
        0 \quad \text{if not}.
    \end{cases}
    \label{pi}
\end{equation}


\subsubsection{Disposal function and probability of purchase}

~

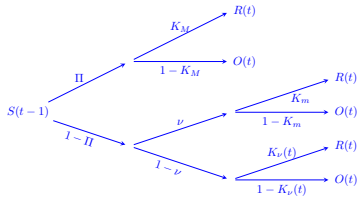
\begin{figure}[!h]
    \centering
    \scalebox{0.45}{\input{factors}}
    \caption{Schematic representing the three pathways to vehicle disposal}
    \label{fig:factors}
\end{figure}

The disposal function $\sigma_{az}(t)$ could be simply defined to be equal to $1$ if the vehicle has been forbidden and to $0$ if it has not been. However, to incorporate realistic behaviour of vehicle owners that do not necessarily give up on their vehicle if it is forbidden or those who dispose of theirs even before it is forbidden, we introduce factors $K_M<1$ and $K_m>0$ that can be valued more accurately through additional studies. 
 Figure~\ref{fig:factors} illustrates the three different pathways of vehicles' disposal considered for our study.
If a vehicle has been forbidden, its owner will dispose of it with a probability of $K_M$. If the vehicle has not been forbidden in zone $z$, two pathways are feasible depending on the number of bans in neighbouring areas. If the vehicle is not forbidden in $z$, nor in $z$'s neighbours, its owner will dispose of the vehicle with a ratio $K_m$. The third pathway treats vehicles that have not been forbidden in their zone $z$ but in neighbouring zones $z'$, where the applied disposal ratio is $K_\nu(t)$.
 Let $\nu(z)$ be the number of $z$'s neighbours and $\nu_z=\mathbf{1}_{\{\nu(z) = \emptyset\}}$. The more neighbouring zones where the vehicle is forbidden, the higher the disposal ratio $K_m < K_\nu(t)<K_M$ will be. 
$K_\nu(t)$ is expressed as
\begin{equation}
    K_\nu (t) = \text{min}\left(K_{lim}\sum_{z'\in \nu(z)} \Pi_{az'}(t) , K_{lim,max}\right).
\end{equation}
By evaluating the number of neighbours of $z$ where the vehicle is forbidden and multiplying it by $K_{lim}$, we scale the impact of restrictive neighbours. To avoid exceeding $K_M$ which is the maximum allowed ratio, we limit the $K_{\nu}(t)$ ratio using $K_{lim, max}$, such that
$K_m < K_{lim}<K_{lim,max}< K_M$. 

Finally, the formulation of the disposal function is 
\begin{equation}
    \sigma_{az}(t) = K_M\Pi_{az}(t) + (\nu_z K_m + (1-\nu_z) K_{\nu}(t))(1-\Pi_{az}(t)). 
    \label{sigma}
\end{equation}


In order to determine the values of both probabilities of purchase (thermal and electric) (cf.~(\ref{probabilities})), we use the utility functions $U_1$ and $U_2$ deployed in \cite{misc-Vinith} (cf. Equation (6)) and a logit discrete choice model (cf. Equation (5) in \cite{misc-Vinith}). We would not allow the purchase of thermal vehicles for zone $z$ if all thermal vehicles have been definitively forbidden for all ages, i.e. $\Pi_{0z}(t) = 1$. Here are the expressions of such ratios
\begin{subequations}
    \begin{align}
        P_{1z}(t) &= \frac{e^{\mu U_1(t)}}{\sum_v e^{\mu U_v (t)}}(1-\Pi_{0z}(t)),
        \label{P1}\\
        P_{2z}(t) &= \frac{e^{\mu U_2(t)}}{\sum_v e^{\mu U_v (t)}}(1-\Pi_{0z}(t)) + \Pi_{0z}(t).
        \label{P2}
\end{align}
\label{probabilities}
\end{subequations}



\subsection{Optimal Control Problem (OCP)}

In order to account for the well-being of citizens and acceptance of the implementation of LEZs, we opted for a constrained optimization framework. The aim is to find the optimal decision variable $J$ that minimizes the total number of vehicles that were disposed of because of the LEZ throughout the years, $R$, under the constraint that CO$_2$ emissions at horizon time $T$ are below a threshold $\Bar{E}$ that can be fixed by decision-makers and that should not be exceeded to respect, e.g., the goal of the European Commission for 2050.

Through~(\ref{new vehicles}), (\ref{old-stock}), (\ref{riddance-stock}), (\ref{new_vehicles_type}), (\ref{stock}), (\ref{relation J et I}), (\ref{pi}), (\ref{sigma}) and (\ref{probabilities}), we express the state variables $S_{vaz}(t)$ and $J_z(t)$ as 
\begin{subequations}
    \begin{align}
        S_{vaz}(t) &= f_S(J_z(t), S_{vaz}(t-1), I_z(t-1), t), \label{S_state}\\ 
        I_z(t) &= f_I(J_z(t), I_z(t-1)). \label{I_state}
    \end{align}
    \label{state}
\end{subequations}
Similarly, through (\ref{riddance-stock}), (\ref{pi}) and (\ref{sigma}), $R_{vaz}(t)$ is expressed as 
\begin{equation}
    \begin{aligned}
        R_{vaz}(t) = f_R(J_z(t), I_z(t-1), S_{vaz}(t-1)).
    \end{aligned}
\end{equation}
Finally, through (\ref{emission_function}) and (\ref{S_state}), the emissions are expressed as
\begin{equation}
    \begin{aligned}
        E(t) = f_E(J_z(t), I_z(t-1), S_{vaz}(t-1), t).
    \end{aligned}
\end{equation}

In summary, the OCP is formulated as follows
\begin{equation}
\begin{array}{|l l}
    \text{Objective function} & \\
     \quad \begin{aligned}\min_{J} R = \sum_{1\leq t\leq T} \sum_{az} f_R(J_z(t), I_z(t-1), S_{1az}(t-1))\end{aligned}  \\[1ex]

    \text{State equations} & \\[0.5ex]
    \quad \biggl(\begin{matrix}
            S_{vaz}(t) \\
            I_z(t)
        \end{matrix}\biggl) = \biggl(\begin{matrix}f_S(J_z(t), S_{vaz}(t-1), I_z(t-1), t) \\
        f_I(J_z(t), I_z(t-1))
        \end{matrix}\biggl) \\[1ex]


    \text{State constraints} & \\[0.5ex]
    \quad S_{vaz}(t) > 0 \\[0.5ex]
    \quad 0 \leq I_z(t) \leq A \\[1ex]

    \text{Control constraint} & \\[0.5ex]
    \quad -1 \leq J_z(t) \leq D_z \\[1ex]

    \text{Terminal constraint} & \\[0.5ex]
    \quad E(T)= f_E(J_z(T), I_z(T-1), S_{vaz}(T-1), T) \leq \Bar{E} \\[1ex]

    \text{Initial conditions} & \\[0.5ex]
    \quad S(0) ~ \text{ (cf. Subsection~\ref{exogenous} and Table~\ref{tab:schedule})} \\[0.5ex]
    \quad I(0)~ \text{ (cf. Table~\ref{tab:schedule})} \\[0.5ex]
\end{array}
\label{ocp}
\end{equation}

The parameter variables seen in Section~\ref{Model_1}, have been expressed as functions of the state variables and do not explicitly appear in the formulation of the problem. Given its complexity, we opted for a direct numerical resolution, using a genetic algorithm (cf. Section~\ref{results}).

\section{Case Study : Île-de-France Region}
\label{case-study}
\begin{figure}[!h]
    \centering
    \includegraphics[width=0.65\linewidth]{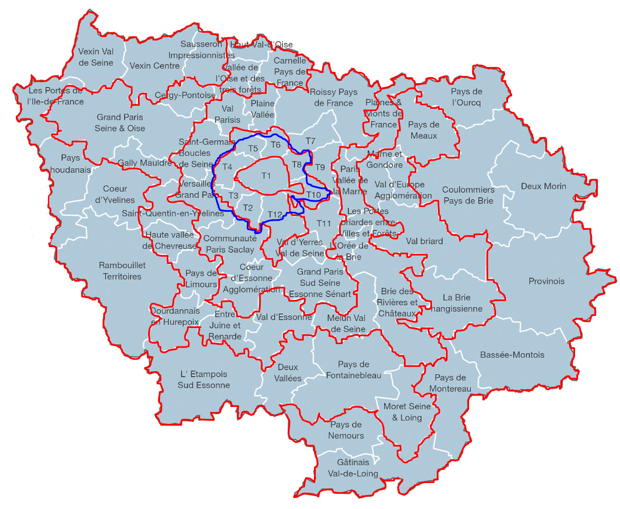}
    \caption{Boundries for the $6$ chosen rigns of the Île-de-France region (red), for the $63$ intermunicipalities (white) and for the current LEZ in the Parisian region (blue)}
    \label{intermunicipalities}
\end{figure}

This paper focuses on the evolution of the vehicle fleet and the implementation of LEZs in the Île-de-France region. The latter is divided into complementary sections where each one would have its own LEZ policy. Many zonings are possible for the Île-de-France region: its 1276 municipalities, its 63 intermunicipalities or its 8 departments for example.
For this study's zoning, we defined successive rings expanding outward from Paris. The first zone corresponds to Paris itself. The second zone includes the intermunicipalities bordering Paris. The third zone contains the intermunicipalities neighbouring those in the second zone, and so on. We obtain 6 rings for the Île-de-France region as shown in Fig~\ref{intermunicipalities}.

\subsection{The initial schedule for 2025}
The Greater Paris Metropolis has implemented LEZs in the Parisian region to improve air quality. It has started implementing LEZs since 2019 where non-classified vehicles (vehicles produced before 1998) were forbidden. The interdictions became more strict over the years as the vehicle fleet is continuously being renewed with safer vehicles which emission levels need to respect fixed European standards known as Euro Norms. 
The latest policy \cite{servicepublic2024}, implemented on January $1^{\text{st}}$ 2025 concerns forbidding diesel cars over 14 years of age and gasoline cars over 19 years of age (i.e. Crit'Air 3 vehicles) in the current Parisian LEZ (LEZ-m) (cf. 
\cite{article-me}). Since the distribution of diesel and gasoline vehicles is quite similar in Île-de-France ($49.7\%$ of vehicles are diesel and $48.2\%$ are gasoline (2021) \cite{data-fleet}), we can consider that the average forbidden vehicle age is 16 or 17 years. This remains an approximation due to the distribution based on whether the vehicle is thermal or electric only. More precision will be possible in future studies where we segment thermal vehicles by fuel type. As for the current Parisian LEZ, it does not precisely align with the rings introduced in this paper, but the closest to its surface would be to combine ring $1$ and ring $2$. 

Table~\ref{tab:schedule} shows the chosen initial values of the state variable $I_z(t)$ and the maximum allowed slope $D_z$ for each ring $z$, both of which we will base the rest of the study on. 
The values $A+1$ for $I_z(0)$ mean that only vehicles of age $A+1$ and more are forbidden. In this study, we consider that the oldest existing vehicles are of age $A$ which implies that forbidding $A+1$ is equivalent to no interdiction in the specified ring.
\begin{table}[!h]
    \centering
    \caption{Initial values of the state variables and the chosen maximum slope $D_z$ for each ring}
    \label{tab:schedule}
    \begin{tabular}{c|cccccc} 
         &  Ring 1&  Ring 2&  Ring 3&  Ring 4&  Ring 5& Ring 6\\ \hline
         $I_z(0)$ &  $16$&  $17$&  A$+1$&  A$+1$&  A$+1$& A$+1$\\ \hline
         $D_z$ &  $4$&  $3$&  $3$&  $2$&  $2$& $1$\\ \hline
         $S_1(0)$ & $1.15$M & $2.09$M & $1.95$M & $0.89$M & $0.43$M & $0.16$M \\ \hline 
         $S_2(0)$ & $8.99$K & $16.31$K & $15.18$K & $6.95$K & $3.37$K & $1.23$K \\ \hline 
    \end{tabular}
\end{table}
\subsection{Parameters and exogenous inputs}
\label{exogenous}
The state variable $S$, was initialized for thermal ($S_1(0)$) and electric ($S_2(0)$) vehicles using french national level stock data from the French Ministry for the ecological Transition and Territorial Cohesion \cite{parc_circulation_vehicles_2025} and adding a segmentation to only account for the specified rings of the Île-de-France region using the right proportion of the population (cf. Table~\ref{tab:schedule}). 
Variables like the survival rate $\eta_a$, the emission factors $\epsilon_{va}(t)$ and $M(t)$ used in~(\ref{emission_function}) are explicitly estimated using other data sources in \cite{misc-Vinith}. The demand $G(t)$, used in~(\ref{new vehicles}) and other exogenous parameters of the utility functions ($U_1$ and $U_2$) and $\mu$ used in the logit model~(\ref{probabilities}) are approximated based on the $\text{DRIVE}^{\text{RS}}$ fleet model of the French Agency of Ecological Transition (ADEME) (cf. \cite{misc-Vinith}).
\subsection{Genetic algorithm}
\label{ga}
We chose a population size of $50$, a number of generations of $1000$, a crossover rate of $0.5$ and a mutation rate of $0.3$ as our genetic algorithm parameters.
\section{Results}
\label{results}

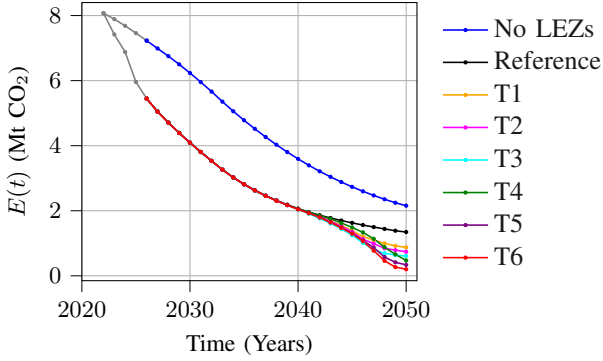
\begin{figure}[ht]
    \centering
    \input{emissions}
    \caption{Annual Emissions from 2026 to 2050}
    \label{fig:emissions_graphe}
\end{figure}

For our backcasting runs, we have chosen a set of different scenarios:
\begin{itemize}
    \item A "No LEZs" scenario where we suppose that LEZs have never existed in the Paris Region.
    \item A "reference" scenario where we suppose that no other new restriction will take place after 2025. It means that what has been forbidden in previous years is still going to be forbidden, but no new interdiction will be introduced after 2025's policy. The emissions for this reference scenario will be referred to as $\Bar{E}_{ref}$.
    \item The optimized scenario showing solutions of OCP~(\ref{ocp}) for six different targets $\Bar{E}^{(i)}$.
    For each target $i$, we want the terminal constraint to verify 
    \begin{equation}
        \Bar{E}^{(i)}(T)\leq \Bar{E}_{ref}(T)(1-\beta^{(i)})
    \end{equation}
    
\end{itemize}
Table~\ref{tab:decrease} shows the decrease percentages $\beta^{(i)}$ for each target.
\begin{table}[ht]
    \centering
    \caption{Desired decrease percentage for each target}
    \label{tab:decrease}
    \begin{tabular}{c|cccccc} 
         Targets & 1 & 2 & 3 & 4 & 5 & 6 \\ \hline
         $\beta^{(i)}$ & 35\% & 45\% & 55\% & 65\% & 75\% & 85\% \\
    \end{tabular}
\end{table}
Figure~\ref{fig:emissions_graphe} represents the emissions obtained for the scenario with no existing LEZs (blue line) in the Île-de-France region, which shape corresponds to the one obtained in \cite{misc-Vinith} 
but for France. Additionally, it shows the emissions obtained for our reference scenario (black line) that was found by evaluating the known reference variable for both $E$ and $R$. The solutions for the six targets were found using the genetic algorithm (cf.~\ref{ga}). Their corresponding emissions are shown in Fig.~\ref{fig:emissions_graphe}.

Most targets of the optimized scenario have the same emissions as the reference scenario from 2026 to 2040, then each one reaches its final value that respects its fixed reduction threshold.

\begin{figure}[ht]
    \centering
    \input{R}
    \caption{Number of vehicles that were disposed of because of the LEZ}
    \label{fig:R_graphe}
\end{figure}
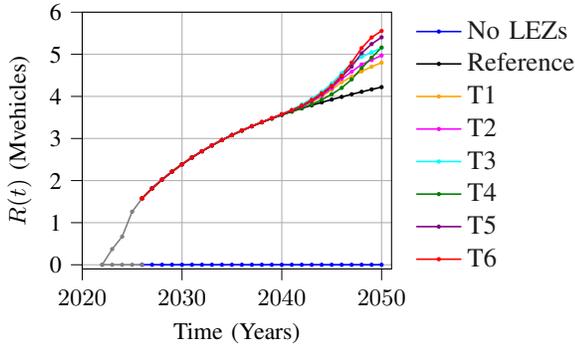

Figure~\ref{fig:R_graphe} shows the cumulative number of vehicles that were disposed of because of the implementation of each LEZ scenario. 
The final value for $T$ (year 2050) corresponds to the running total of $R(t)$ throughout the years and represents the minimization criterion.
In the case of no introduction of LEZs in the Île-de-France region, no vehicle will be disposed of (i.e. $R(t)=0 \text{ } \forall t$ for the blue line). The reference scenario (black line) shows the lowest possible values for $R(t)$ but is not null as people continue to get rid of their vehicles, years after they have been banned. The other genetic algorithm scenario targets show that the higher $R$ is, the lower $E(T)$ will be.

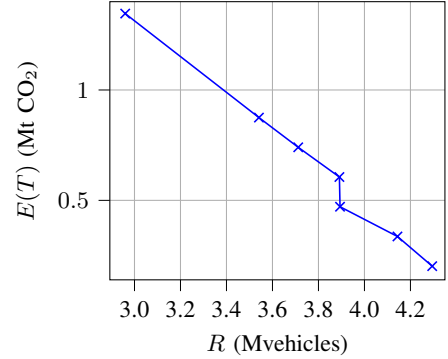
\begin{figure}[ht]
    \centering
    \input{pareto}
    \caption{The Pareto front showing final emissions as a function of the total number of disposed vehicles, for the reference and the optimized scenarios}
    \label{fig:pareto_front_graphe}
\end{figure}

 In constrained optimization, visualizing the Pareto front is essential  for decision-making as it highlights the trade-offs between feasibility and optimality. Decision-makers can make a more informed decision by choosing to relax or tighten constraints and see their impact on the objective function. The Pareto front for the reference scenario and the six targets is shown in Fig.~\ref{fig:pareto_front_graphe}, where $E(T)$ is a function of the objective $R$. 

The highest emissions which correspond to the the lowest $R$ are reached by the reference scenario (black line in Figs.~\ref{fig:emissions_graphe} and~\ref{fig:R_graphe}). The lowest emissions which correspond to the highest $R$ are reached with target 6 (red line in Figs.~\ref{fig:emissions_graphe} and~\ref{fig:R_graphe}) where a reduction of $85\%$ of $\Bar{E}_{ref}(T)$ emissions was targeted.
The two breaking points in Fig.~\ref{fig:pareto_front_graphe} correspond to targets 3 (light blue) and 4 (green). Both seem to have approximately the same $R$ (cf. Figs.~\ref{fig:R_graphe} and~\ref{fig:pareto_front_graphe}) with a difference of $0.7\%$ but target 4 has lower final emissions than target 3 (cf. Figs.~\ref{fig:emissions_graphe} and~\ref{fig:pareto_front_graphe}) with a $22\%$ decrease in emissions compared to target 3. 

\begin{figure}[ht]
    \centering
    \input{decision}
    \caption{Policy schedule suggested by the decision variable of scenario 4}
    \label{fig:decision_variable_graph}
\end{figure}
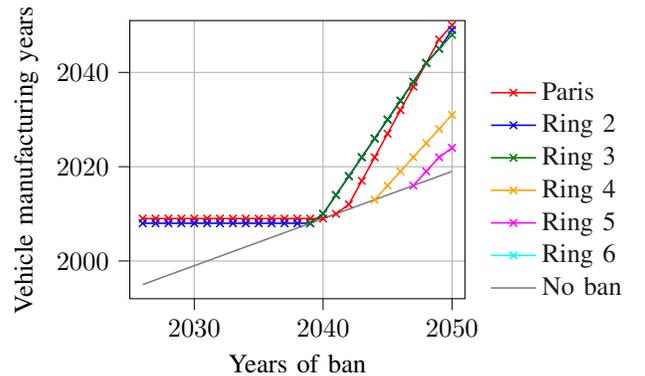

Figure~\ref{fig:decision_variable_graph} represents the policy of target 4, showing the different schedules throughout the years for the six chosen rings. The years of vehicles production going from $1995$ to $2050$ are presented as a function of the year of ban.
According to this policy, no restriction will be applied in ring 6 while restrictions will not be implemented in ring 5 until $2048$, ring 4 until $2045$ and ring 3 until $2040$. No new restrictions are scheduled between $2026$ and $2040$ for rings 1 and between $2026$ and $2039$ for ring 2. The restriction policies for ring 2 and ring 3 are at their highest slope : $D_2 = D_3 = 3$ from $2040$ to $2048$. Ring 1 takes more time to start its highest slope $D_1 = 4$ from $2042$ to $2049$ but still reaches the highest point for $2050$. 
Generally, we can say that the closer to ring 1 (Paris) the stricter the restrictions are. With this policy, we obtain a total number of vehicle disposal of $R = 5.2~10^6$, which represents $78\%$ of $S_1(0)$, and corresponding final emissions valued at $E(T) = 471$ ktCO$_2$, showing a decrease of $92\%$ compared to the initial emission value $E(0)$ of year $2025$. 

\section{Conclusion and perspectives}
\label{conclusion}
Applying a backcasting framework allows policymakers to ask the right questions, prioritize key targets, and explore innovative strategies to address mobility challenges. By continuously refining methodologies and integrating diverse policy levers, we can move towards more sustainable and efficient urban transport systems that benefit both the environment and public health. This paper has proven that the backcasting approach is feasible. Applied to LEZ policies in the Île-de-France region, and resolved using a genetic algorithm, the run results suggest the proper schedule of LEZ implementation to reach chosen emission targets, while limiting impacts on individuals' mobility by minimizing the number of disposed vehicles in response to the introduction of LEZs.

Future research can further enhance the backcasting approach in several ways. A finer disaggregation than the rings introduced in this paper could be explored to better reflect local characteristics in the implementation of LEZs. 
Incorporating thermal vehicle types, such as diesel, gasoline and plug-in hybrid vehicles, would provide a more accurate representation of the vehicle fleet and its evolution. User behaviour leading to the disposal ratio $\sigma$ could be better investigated and modeled with additional studies. 
Furthermore, while the primary objective remains the reduction of greenhouse gas emissions, it would be interesting for future research to account for air quality indicators such as nitrogen oxides (NO$_x$) and particulate matter (PM). 
Combining LEZs with other measures, such as incentives for electric vehicle adoption, improvements in public transport infrastructure, or urban planning strategies that encourage active mobility (e.g., cycling infrastructure and safety measures), could lead to more effective policy roadmaps; a diversified approach would not only optimize environmental benefits but also enhance economic feasibility and social acceptance. 
The genetic algorithm has provided initial results, but further validation is needed. Other methods such as parametric optimization ought to be explored and applied to better solve the problem. 
\section*{Acknowledgment}
This research benefited from aid managed by the \textit{Agence nationale de la recherche (ANR)}, under France 2030, within the project FORBAC bearing the reference ANR-23-PEMO-0002.

\bibliographystyle{IEEEtran}
\input{main.bbl}

\end{document}

%% file: survival.tex
\begin{tikzpicture}[>=stealth, node distance=2cm, scale=0.7, transform shape, every node/.style={align=center}]

    \node (S) at (0,0) {\textbf{\textcolor{blue}{$S(t-1)$}}};
    \node (mid) at (3,0.5) {};
    \node (R) at (6,1) {\textbf{\textcolor{blue}{$R(t)$}}};
    \node (O) at (6,0) {\textbf{\textcolor{blue}{$O(t)$}}};

    \draw[->, thick, blue] (S) -- (mid) node[midway, above] {$\eta$};
    \draw[->, thick, blue] (mid) -- (R) node[midway, above] {$\sigma$};
    \draw[->, thick, blue] (mid) -- (O) node[midway, below] {$1-\sigma$};
    \draw[->, thick, blue] (S) -- ++(3,-0.5) node[midway, below] {$1-\eta$};

\end{tikzpicture}

%% file: factors.tex
\usetikzlibrary{positioning}
\begin{tikzpicture}[>=stealth, node distance=1.5cm and 2cm, 
transform shape, every node/.style={align=center}]

    \node (S) at (0,0) {\textbf{\textcolor{blue}{$S(t-1)$}}};

    \node (Pi) at (3,1.5) {};
    \node (OnePi) at (3,-1) {};

    \node (KM) at (6,3) {};
    \node (OneKM) at (6,1.5) {};
    \node (Nu) at (6,0) {};
    \node (OneNu) at (6,-2) {};

    \node (R1) at (6.4,3) {\textbf{\textcolor{blue}{$R(t)$}}};
    \node (O1) at (6.4,1.5) {\textbf{\textcolor{blue}{$O(t)$}}};
    \node (Km) at (9,1) {};
    \node (OneKm) at (9,0) {};
    \node (Knu) at (9,-1) {};
    \node (OneKnu) at (9,-2) {};

    \node (R2) at (9.4,1) {\textbf{\textcolor{blue}{$R(t)$}}};
    \node (O2) at (9.4,0) {\textbf{\textcolor{blue}{$O(t)$}}};
    \node (R3) at (9.4,-1) {\textbf{\textcolor{blue}{$R(t)$}}};
    \node (O3) at (9.4,-2) {\textbf{\textcolor{blue}{$O(t)$}}};

    \draw[->, thick, blue] (S) -- (Pi) node[pos=0.4, above] {$\Pi$};
    \draw[->, thick, blue] (S) -- (OnePi) node[pos=0.4, below, sloped] {$1-\Pi$};

    \draw[->, thick, blue] (Pi) -- (KM) node[midway, above] {$K_M$};
    \draw[->, thick, blue] (Pi) -- (OneKM) node[midway, below] {$1-K_M$};

    \draw[->, thick, blue] (OnePi) -- (Nu) node[midway, above] {$\nu$};
    \draw[->, thick, blue] (OnePi) -- (OneNu) node[pos=0.4, below, sloped] {$1-\nu$};


    \draw[->, thick, blue] (Nu) -- (Km) node[pos=0.7, below] {$K_m$};
    \draw[->, thick, blue] (Nu) -- (OneKm) node[midway, below] {$1-K_m$};

    \draw[->, thick, blue] (OneNu) -- (Knu) node[midway, above] {$K_{\nu}(t)$};
    \draw[->, thick, blue] (OneNu) -- (OneKnu) node[midway, below] {$1-K_{\nu}(t)$};


\end{tikzpicture}

%% file: emissions.tex
\begin{tikzpicture}

\definecolor{cyan}{RGB}{0,255,255}
\definecolor{darkgray176}{RGB}{176,176,176}
\definecolor{green}{RGB}{0,128,0}
\definecolor{magenta}{RGB}{255,0,255}
\definecolor{purple}{RGB}{128,0,128}
\definecolor{orange}{RGB}{255,165,0}

\begin{axis}[
tick align=outside,
tick pos=left,
x grid style={darkgray176},
scale = 0.65,
xlabel={Time (Years)},
xmajorgrids,
xmin=2020, xmax=2051, 
xtick style={color=black},
xticklabel style={/pgf/number format/set thousands separator={}},
y grid style={darkgray176},
ylabel={$E(t)$ (Mt CO$_2$)},
ymajorgrids,
ymin=-0.149410155564568, ymax=8.4,
ytick style={color=black},
legend style={at={(1.05,0.98)}, 
anchor=north west, 
legend cell align=left, 
align=left, 
draw=none, 
fill opacity=0.9},
xlabel style={font=\small}, 
ylabel style={font=\small}, 
xticklabel style={font=\small}, 
yticklabel style={font=\small},
]
\addplot [semithick, blue, mark=*, mark size=.5, mark options={solid}]
table {%
2026 7.22833554799088
2027 6.98876802826812
2028 6.75310979914955
2029 6.50054256413386
2030 6.23469103284543
2031 5.95486882049962
2032 5.6632148345306
2033 5.3537036416895
2034 5.06031380217245
2035 4.78248030730553
2036 4.5188867200296
2037 4.26680555056669
2038 4.02876498809163
2039 3.80471729726795
2040 3.59426752392461
2041 3.3972579474902
2042 3.21319814651379
2043 3.04184930276496
2044 2.88256494829563
2045 2.7348083412944
2046 2.59794927186413
2047 2.47126550371892
2048 2.35386222419471
2049 2.2449430728281
2050 2.15608055723446
};
\addlegendentry{No LEZs}

\addplot [semithick, black, mark=*, mark size=.5, mark options={solid}]
table {%
2026 5.4440628229424
2027 5.04724773014908
2028 4.70826096125901
2029 4.3910724523462
2030 4.09179485170726
2031 3.80588796447584
2032 3.53192771082677
2033 3.26255748077521
2034 3.02467643997515
2035 2.81441545431588
2036 2.62818235077169
2037 2.46173663210312
2038 2.31394198606891
2039 2.18246982067259
2040 2.06484376072245
2041 1.95899210883792
2042 1.86370523234525
2043 1.77756767582982
2044 1.69917148458358
2045 1.62725942835229
2046 1.56079342291818
2047 1.49890852853492
2048 1.44085868387648
2049 1.38609233992754
2050 1.34607785246176
};
\addlegendentry{Reference}

\addplot [semithick, orange, mark=*, mark size=.5, mark options={solid}]
table {%
2026 5.4440628229424
2027 5.04724773014908
2028 4.70826096125901
2029 4.3910724523462
2030 4.09179485170726
2031 3.80588796447584
2032 3.53192771082677
2033 3.26255748077521
2034 3.02467643997515
2035 2.81441545431588
2036 2.62818235077169
2037 2.46166492302524
2038 2.31318449790823
2039 2.17943146722661
2040 2.05701525911943
2041 1.94184035269658
2042 1.81977727089929
2043 1.69188829346916
2044 1.55208419499303
2045 1.38331049445571
2046 1.2115297882081
2047 1.09774388264081
2048 0.997272707920757
2049 0.920912910183298
2050 0.874950290801357
};
\addlegendentry{T1}

\addplot [semithick, magenta, mark=*, mark size=.5, mark options={solid}]
table {%
2026 5.4440628229424
2027 5.04724773014908
2028 4.70826096125901
2029 4.3910724523462
2030 4.09179485170726
2031 3.80588796447584
2032 3.53192771082677
2033 3.26255748077521
2034 3.02467643997515
2035 2.81441545431588
2036 2.62818235077169
2037 2.46173663210312
2038 2.31394198606891
2039 2.17885910112877
2040 2.05466468376959
2041 1.93647998198911
2042 1.81281338640073
2043 1.67259307659537
2044 1.52178459980334
2045 1.34674691352954
2046 1.12642140080968
2047 0.988417015797768
2048 0.847141534011502
2049 0.789428181720887
2050 0.740342347498927
};
\addlegendentry{T2}

\addplot [semithick, cyan, mark=*, mark size=.5, mark options={solid}]
table {%
2026 5.4440628229424
2027 5.04724773014908
2028 4.70826096125901
2029 4.3910724523462
2030 4.09179485170726
2031 3.80588796447584
2032 3.53192771082677
2033 3.26255748077521
2034 3.02467643997515
2035 2.81432702283645
2036 2.62794856510376
2037 2.46148135551021
2038 2.30990541901822
2039 2.17047347398004
2040 2.04004691409803
2041 1.90971538378212
2042 1.76565371676168
2043 1.6108905438841
2044 1.4450523801378
2045 1.2486217202276
2046 1.01452045465411
2047 0.79392859563806
2048 0.693935568151105
2049 0.638340224896911
2050 0.605734943311612
};
\addlegendentry{T3}
\addplot [semithick, green, mark=*, mark size=.5, mark options={solid}]
table {%
2026 5.4440628229424
2027 5.04724773014908
2028 4.70826096125901
2029 4.3910724523462
2030 4.09179485170726
2031 3.80588796447584
2032 3.53192771082677
2033 3.26255748077521
2034 3.02467643997515
2035 2.81441545431588
2036 2.62818235077169
2037 2.46173663210312
2038 2.31394198606891
2039 2.18246982067259
2040 2.0638450983222
2041 1.9541067137556
2042 1.85089492932669
2043 1.75304912288216
2044 1.6241730553143
2045 1.48684454581254
2046 1.33543178955935
2047 1.13716405691182
2048 0.880034760655587
2049 0.66086959298202
2050 0.471126996283287
};
\addlegendentry{T4}

\addplot [semithick, purple, mark=*, mark size=.5, mark options={solid}]
table {%
2026 5.4440628229424
2027 5.04724773014908
2028 4.70826096125901
2029 4.3910724523462
2030 4.09179485170726
2031 3.80588796447584
2032 3.53192771082677
2033 3.26255748077521
2034 3.02467643997515
2035 2.81441545431588
2036 2.62818235077169
2037 2.46173663210312
2038 2.31307596700942
2039 2.1771514696359
2040 2.05117683664944
2041 1.93091270187324
2042 1.80415679408973
2043 1.65206635297647
2044 1.48861837535506
2045 1.30812074905561
2046 1.08590742983872
2047 0.86910335395567
2048 0.578183017455668
2049 0.415593004729296
2050 0.336432801035623
};
\addlegendentry{T5}

\addplot [semithick, red, mark=*, mark size=.5, mark options={solid}]
table {%
2026 5.4440628229424
2027 5.04724773014908
2028 4.70826096125901
2029 4.3910724523462
2030 4.09179485170726
2031 3.80588796447584
2032 3.53192771082677
2033 3.26255748077521
2034 3.02467643997515
2035 2.81418132399057
2036 2.62643545324242
2037 2.45733277497553
2038 2.30619422419373
2039 2.16979855186476
2040 2.04231659442689
2041 1.92228776558635
2042 1.79352106947118
2043 1.64000384042253
2044 1.47473870338286
2045 1.2947022719389
2046 1.0663870579051
2047 0.773867270809251
2048 0.462201919187253
2049 0.26699614724335
2050 0.201911068414263
};
\addlegendentry{T6}

\addplot [semithick, gray, mark=*, mark size=.5, mark options={solid}]
table {%
2022 8.07231098680322
2023 7.88922460082881
2024 7.68485813917704
2025 7.46038206742696
2026 7.22833554799088
};

\addplot [semithick, gray, mark=*, mark size=.5, mark options={solid}]
table {%
2022 8.07231098680322
2023 7.42100225278879
2024 6.88082926793071
2025 5.95741126806475
2026 5.4440628229424
};

\addplot [semithick, red, mark=*, mark size=.5, mark options={solid}]
table {%
2026 5.4440628229424
};

\addplot [semithick, blue, mark=*, mark size=.5, mark options={solid}]
table {%
2026 7.22833554799088
};

\end{axis}

\end{tikzpicture}

%% file: R.tex
\begin{tikzpicture}

\definecolor{cyan}{RGB}{0,255,255}
\definecolor{darkgray176}{RGB}{176,176,176}
\definecolor{green}{RGB}{0,128,0}
\definecolor{magenta}{RGB}{255,0,255}
\definecolor{purple}{RGB}{128,0,128}
\definecolor{orange}{RGB}{255,165,0}

\begin{axis}[
tick align=outside,
scale = 0.6,
tick pos=left,
x grid style={darkgray176},
xlabel={Time (Years)},
xmajorgrids,
xmin=2020, xmax=2051,
xtick style={color=black},
xticklabel style={/pgf/number format/set thousands separator={}},
y grid style={darkgray176},
ylabel={$R(t)$ (Mvehicles)},
ymajorgrids,
ymin=-0.1, ymax=6,
ytick style={color=black},
ytick={0,1,2,3,4,5,6},
yticklabels={0,1,2,3,4,5,6},
xlabel style={font=\small}, 
ylabel style={font=\small}, 
xticklabel style={font=\small}, 
yticklabel style={font=\small},
legend style={at={(1.05,1.01)}, 
anchor=north west, 
legend cell align=left, 
align=left, 
draw=none, 
fill opacity=0.9},
]

\addplot [semithick, blue, mark=*, mark size=.5, mark options={solid}]
table {%
2026 0
2027 0
2028 0
2029 0
2030 0
2031 0
2032 0
2033 0
2034 0
2035 0
2036 0
2037 0
2038 0
2039 0
2040 0
2041 0
2042 0
2043 0
2044 0
2045 0
2046 0
2047 0
2048 0
2049 0
2050 0
};
\addlegendentry{No LEZs}

\addplot [semithick, black, mark=*, mark size=.5, mark options={solid}]
table {%
2026 1.575774
2027 1.814420
2028 2.022398
2029 2.211802
2030 2.386457
2031 2.547964
2032 2.697331
2033 2.835234
2034 2.962092
2035 3.079328
2036 3.188183
2037 3.289712
2038 3.384815
2039 3.474267
2040 3.558930
2041 3.639477
2042 3.716032
2043 3.788946
2044 3.858518
2045 3.925076
2046 3.988887
2047 4.050170
2048 4.109090
2049 4.165775
2050 4.220348
};
\addlegendentry{Reference}

\addplot [semithick, orange, mark=*, mark size=.5, mark options={solid}]
table {%
2026 1.575774
2027 1.814420
2028 2.022398
2029 2.211802
2030 2.386457
2031 2.547964
2032 2.697331
2033 2.835234
2034 2.962092
2035 3.079328
2036 3.188183
2037 3.289761
2038 3.385349
2039 3.476526
2040 3.565141
2041 3.653911
2042 3.753946
2043 3.866173
2044 3.996827
2045 4.160861
2046 4.336236
2047 4.468406
2048 4.594056
2049 4.704615
2050 4.801972
};
\addlegendentry{T1}

\addplot [semithick, magenta, mark=*, mark size=.5, mark options={solid}]
table {%
2026 1.575774
2027 1.814420
2028 2.022398
2029 2.211802
2030 2.386457
2031 2.547964
2032 2.697331
2033 2.835234
2034 2.962092
2035 3.079328
2036 3.188183
2037 3.289712
2038 3.384815
2039 3.476807
2040 3.566801
2041 3.658163
2042 3.760003
2043 3.882582
2044 4.024242
2045 4.197679
2046 4.421916
2047 4.587246
2048 4.759151
2049 4.863472
2050 4.972809
};
\addlegendentry{T2}

\addplot [semithick, cyan, mark=*, mark size=.5, mark options={solid}]
table {%
2026 1.575774
2027 1.814420
2028 2.022398
2029 2.211802
2030 2.386457
2031 2.547964
2032 2.697331
2033 2.835234
2034 2.962092
2035 3.079388
2036 3.188358
2037 3.289940
2038 3.387756
2039 3.483598
2040 3.579652
2041 3.682706
2042 3.804564
2043 3.945205
2044 4.106686
2045 4.306167
2046 4.550270
2047 4.794100
2048 4.939096
2049 5.049964
2050 5.151865
};
\addlegendentry{T3}

\addplot [semithick, green, mark=*, mark size=.5, mark options={solid}]
table {%
2026 1.575774
2027 1.814420
2028 2.022398
2029 2.211802
2030 2.386457
2031 2.547964
2032 2.697331
2033 2.835234
2034 2.962092
2035 3.079328
2036 3.188183
2037 3.289712
2038 3.384815
2039 3.474267
2040 3.559614
2041 3.643070
2042 3.726207
2043 3.809898
2044 3.922016
2045 4.050721
2046 4.201898
2047 4.404296
2048 4.671232
2049 4.919768
2050 5.161881
};
\addlegendentry{T4}

\addplot [semithick, purple, mark=*, mark size=.5, mark options={solid}]
table {%
2026 1.575774
2027 1.814420
2028 2.022398
2029 2.211802
2030 2.386457
2031 2.547964
2032 2.697331
2033 2.835234
2034 2.962092
2035 3.079328
2036 3.188183
2037 3.289712
2038 3.385412
2039 3.478144
2040 3.569692
2041 3.663146
2042 3.768103
2043 3.901020
2044 4.054949
2045 4.236381
2046 4.467114
2047 4.707656
2048 5.025998
2049 5.244204
2050 5.404303
};
\addlegendentry{T5}

\addplot [semithick, red, mark=*, mark size=.5, mark options={solid}]
table {%
2026 1.575774
2027 1.814420
2028 2.022398
2029 2.211802
2030 2.386457
2031 2.547964
2032 2.697331
2033 2.835234
2034 2.962092
2035 3.079488
2036 3.189432
2037 3.293136
2038 3.391366
2039 3.485615
2040 3.579440
2041 3.673955
2042 3.781614
2043 3.916752
2044 4.073389
2045 4.256041
2046 4.494180
2047 4.803539
2048 5.147109
2049 5.400841
2050 5.555433
};
\addlegendentry{T6}

\addplot [semithick, gray, mark=*, mark size=.5, mark options={solid}]
table {%
2022 0
2023 0
2024 0
2025 0
2026 0
};

\addplot [semithick, gray, mark=*, mark size=.5, mark options={solid}]
table {%
2022 0
2023 0.371191
2024 0.668620
2025 1.260766
2026 1.575774
};

\addplot [semithick, red, mark=*, mark size=.5, mark options={solid}]
table {%
2026 1.575774
};

\end{axis}

\end{tikzpicture}

%% file: pareto.tex
\begin{tikzpicture}

\definecolor{darkgray176}{RGB}{176,176,176}

\begin{axis}[
tick align=outside,
tick pos=left,
scale = 0.65,
x grid style={darkgray176},
xlabel={$R$ (Mvehicles)},
xmajorgrids,
xmin=2.89282766704142, xmax=4.35,
xtick style={color=black},
xtick={3,3.2,3.4,3.6,3.8,4,4.2,4.4},
xticklabels={3.0,3.2,3.4,3.6,3.8,4.0,4.2,4.4},
y grid style={darkgray176},
ylabel={$E(T)$ (Mt CO$_2$)},
ymajorgrids,
ymin=0.14, ymax=1.4,
ytick style={color=black},
xlabel style={font=\small}, 
ylabel style={font=\small}, 
xticklabel style={font=\small}, 
yticklabel style={font=\small},
]
\addplot [semithick, blue, mark=x, mark size=2.5, mark options={solid}]
table {%
2.9595819299109 1.34607785246176
3.54120603034733 0.874950290801357
3.71204241596395 0.740342347498927
3.8910990963667 0.605734943311612
3.89381631470313 0.471126996283287
4.14353710314185 0.336432801035623
4.29466718730053 0.201911068414263
};
\end{axis}

\end{tikzpicture}

%% file: decision.tex
\begin{tikzpicture}

\definecolor{cyan}{RGB}{0,255,255}
\definecolor{darkgray176}{RGB}{176,176,176}
\definecolor{green}{RGB}{0,128,0}
\definecolor{magenta}{RGB}{255,0,255}
\definecolor{orange}{RGB}{255,165,0}

\begin{axis}[
tick align=outside,
scale = 0.65,
tick pos=left,
x grid style={darkgray176},
xlabel={Years of ban},
xmajorgrids,
xmin=2025, xmax=2051,
xtick style={color=black},
xticklabel style={/pgf/number format/set thousands separator={}},
y grid style={darkgray176},
ylabel={Vehicle manufacturing years},
ymajorgrids,
ymin=1992, ymax=2051,
ytick style={color=black},
yticklabel style={/pgf/number format/set thousands separator={}},
legend style={at={(1.05,0.82)}, 
anchor=north west, 
legend cell align=left, 
align=left, 
draw=none, 
fill opacity=0.9}
]
\addplot [semithick, red, mark=x, mark size=2, mark options={solid}]
table {%
2026 2009
2027 2009
2028 2009
2029 2009
2030 2009
2031 2009
2032 2009
2033 2009
2034 2009
2035 2009
2036 2009
2037 2009
2038 2009
2039 2009
2040 2009
2041 2010
2042 2012
2043 2017
2044 2022
2045 2027
2046 2032
2047 2037
2048 2042
2049 2047
2050 2050
};
\addlegendentry{ Paris}

\addplot [semithick, blue, mark=x, mark size=2, mark options={solid}]
table {%
2026 2008
2027 2008
2028 2008
2029 2008
2030 2008
2031 2008
2032 2008
2033 2008
2034 2008
2035 2008
2036 2008
2037 2008
2038 2008
2039 2008
2040 2010
2041 2014
2042 2018
2043 2022
2044 2026
2045 2030
2046 2034
2047 2038
2048 2042
2049 2045
2050 2049
};
\addlegendentry{ Ring 2}

\addplot [semithick, green, mark=x, mark size=2, mark options={solid}]
table {%
2039 2008
2040 2010
2041 2014
2042 2018
2043 2022
2044 2026
2045 2030
2046 2034
2047 2038
2048 2042
2049 2045
2050 2048
};
\addlegendentry{ Ring 3}

\addplot [semithick, orange, mark=x, mark size=2, mark options={solid}]
table {%
2044 2013
2045 2016
2046 2019
2047 2022
2048 2025
2049 2028
2050 2031
};
\addlegendentry{ Ring 4}

\addplot [semithick, magenta, mark=x, mark size=2, mark options={solid}]
table {%
2047 2016
2048 2019
2049 2022
2050 2024
};
\addlegendentry{ Ring 5}

\addplot [semithick, cyan, mark=x, mark size=2, mark options={solid}]
table {%
2020  1990
};
\addlegendentry{ Ring 6}

\addplot [semithick, gray, mark size=2, mark options={solid}]
table {%
2026 1995
2027 1996
2028 1997
2029 1998
2030 1999
2031 2000
2032 2001
2033 2002
2034 2003
2035 2004
2036 2005
2037 2006
2038 2007
2039 2008
2040 2009
2041 2010
2042 2011
2043 2012
2044 2013
2045 2014
2046 2015
2047 2016
2048 2017
2049 2018
2050 2019
};
\addlegendentry{ No ban}

\end{axis}

\end{tikzpicture}